\documentclass[11pt]{article}
\usepackage{amsmath}
\usepackage{amssymb}
\usepackage{amsfonts}
\usepackage{amsthm}
\usepackage{fullpage}
\usepackage{graphicx}
\usepackage{times}

\DeclareMathAlphabet{\varmathbb}{U}{bbold}{m}{n}
\newcommand{\one}{\varmathbb 1}
\renewcommand{\vec}[1]{\mathbf{#1}}

\begin{document}

\newtheorem{theorem}{Theorem}
\newtheorem{lemma}[theorem]{Lemma}
\newtheorem{conjecture}{Conjecture}
\newtheorem{claim}[theorem]{Claim}
\newtheorem{corollary}[theorem]{Corollary}
\newtheorem{definition}{Definition}
\newtheorem{assumption}{Assumption}

\newcommand{\Exp}{{\rm E}}
\newcommand{\F}{{\mathbb F}}
\newcommand{\GL}{{\rm GL}}
\newcommand{\End}{{\rm End}}
\newcommand{\e}{{\rm e}}
\newcommand{\tr}{\textbf{tr}}
\newcommand{\rk}{\textbf{rk}}

\newcommand{\Z}{\mathbb{Z}}
\newcommand{\R}{\mathbb{R}}

\newcommand{\cons}{\textrm{Cons}}
\newcommand{\spans}{\textrm{Spans}}
\newcommand{\poly}{\textrm{poly}}

\newcommand{\va}{\vec{a}}
\newcommand{\vb}{\vec{b}}
\newcommand{\vx}{\vec{x}}
\newcommand{\vy}{\vec{y}}
\newcommand{\vv}{\vec{v}}
\newcommand{\vg}{\vec{g}}
\newcommand{\vw}{\vec{w}}

\newcommand{\remove}[1]{}

\title{A classical one-way function to confound quantum adversaries}

\author{Cristopher Moore\\ \texttt{moore@cs.unm.edu}\\ University of New Mexico 
\\ and the Santa Fe Institute \and Alexander Russell\\ \texttt{acr@cse.uconn.edu}\\ University of Connecticut \and Umesh Vazirani\\ \texttt{vazirani@cs.berkeley.edu}\\ U. C. Berkeley}

\maketitle

\abstract{The promise of quantum computation and its consequences  
for complexity-theoretic cryptography motivates an
immediate search for cryptosystems which can be implemented with
current technology, but which remain secure even in the presence of
quantum computers. Inspired by recent negative results pertaining to
the nonabelian hidden subgroup problem, we present here a classical
algebraic function $f_V(M)$ of a matrix $M$ which we believe is a 
one-way function 
secure against quantum attacks.  Specifically, inverting
$f_V$ reduces naturally to solving a hidden subgroup problem over the
general linear group (which is at least as hard as the hidden subgroup problem 
over the symmetric group).  We also demonstrate a reduction from Graph
Isomorphism to the problem of inverting $f_V$;  
unlike Graph Isomorphism, however, the function $f_V$ is
random self-reducible and therefore uniformly hard.  

These results suggest that, unlike Shor's algorithm for the discrete logarithm---which 
is, so far, the only successful quantum attack on a classical one-way function---quantum attacks 
based on the hidden subgroup problem are unlikely to work.  
We also show that reconstructing any entry of $M$, or the trace of $M$, with
nonnegligible advantage is essentially as hard as inverting $f_V$.  
Finally, $f_V$ can be efficiently computed and the number of output
bits is less than $1+\epsilon$ times the number of input bits for any
$\epsilon > 0$.
\newpage
\pagenumbering{arabic}
\section{Introduction}

When a quantum computer is finally built, perhaps its most important
practical impact will be on modern cryptography, thanks to Shor's
celebrated quantum algorithms for factoring and discrete
logs~\cite{Shor:1997:PTA} (and a sequence of followup results).
Quantum cryptography provides a partial recourse, though its scope is
limited by ``no-go'' theorems such as the impossibility of quantum bit
commitment, as well as extravagant physical infrastructure
requirements.  A plausible route to a more acceptable antidote was
suggested in a result contemporaneous with Shor's paper, showing that
quantum computers require exponential time to invert a random
permutation in a black box model~\cite{BBBV:1997:BBBV}. Since a random
permutation is a standard abstraction for a one-way function, this
result suggested the possibility of creating classical cryptography
that is resistant to quantum cryptanalysis. The practical challenge is
to design a function $f: \Sigma^n \rightarrow \Sigma^m$ that can be computed 
very efficiently by a classical computer, while providing 
credible evidence that inversion is difficult even 
with a quantum computer. It is also desirable that $f$ be nonexpansive, 
i.e., that $m$ not be much larger than $n$.  This is the goal of this paper.

Our task is facilitated by new insights obtained over the last
few years into the limits of quantum algorithms for 
the non-abelian hidden subgroup problem (HSP).
A series of negative
results~\cite{Hallgren:2000:NSR,Grigni:2001:QMA,Moore:2005:SGF}
culminating in Hallgren, et
al.~\cite{Hallgren:2006:LQC} shows
that for sufficiently non-abelian groups the HSP 
is hard for quantum computers in the sense that any quantum
algorithm using the coset state framework requires exponential
time unless it makes highly entangled measurements of 
$\Omega(\log |G|)$ registers.  Very few algorithmic models for 
highly-entangled measurements are known; one 
of the few proposals for carrying out such measurements efficiently is
a ``quantum sieve,'' developed by Kuperberg~\cite{K} for the HSP on the
dihedral group.  However, a recent result of Moore, Russell, and
\'Sniady~\cite{MRS:Impossibility} shows that no such approach yields an efficient algorithm
over the symmetric groups. In fact, for the cases relevant to Graph Isomorphism, algorithms of this form cannot even do much better than the best known
classical algorithms. This forms the basis of our main assumption about the
limitations of quantum algorithms.

Our function, which we denote $f_V$, is parametrized by a list of
vectors $V={\vec{v}_1,\vec{v}_2,\ldots,\vec{v}_m}$; we will choose each $\vec{v}_i$ independently and uniformly
at random from $\F_q^n$, where $q$ is some small prime.  Then given $M
\in \GL_n(\F_q)$, that is, an invertible $n \times n$ matrix over $\F_q$,
we define $f_V(M)$ as the collection
\[
MV = \{ M\vec{v} \mid \vec{v} \in V\}\enspace.
\]
However, $f_V$ returns this collection as an unordered set (say,
sorted in lexicographic order).  In other words, we know that each $\vec{w}
\in f_V(M)$ is $M\vec{v}$ for some $\vec{v} \in V$, but we do not know with what
permutation the $\vec{v}$s and $\vec{w}$s correspond.

In Section~\ref{sec:injective}, we show that $f_V$ is one-to-one with
high probability in $V$ whenever $m$ is slightly larger than $n$, say $m = n +
O(\ln^2 n)$. Also, clearly $f_V$
can be computed very efficiently, in time $M(n)$, the time to multiply
two $n \times n$ matrices.  As a function of the input length $k =
n^2$, the time is essentially $\sqrt{M(k)}$.

In Section~\ref{sec:hardness}, we point out that the natural reduction of inverting $f_V$ to a
hidden subgroup (or hidden shift) problem results in hidden subgroup
problems on the general linear group $\GL_n$.  This group contains 
the symmetric group $S_n$ as a subgroup, and its HSP appears resistant 
to all known quantum techniques.  Moreover, we reduce the Graph 
Isomorphism problem to the problem of inverting $f_V$.  This 
implies that no quantum attack analogous to Shor's algorithm for 
the discrete logarithm can succeed, unless there is an efficient 
quantum algorithm for Graph Isomorphism. 

We stress that unlike Graph Isomorphism, for which there is no known way to generate 
hard random instances, inverting $f_V$ is uniformly hard because of the following simple 
observation: for any matrix $A$, we have $f_V(AM) = A f_V(M)$.  By choosing $A$ randomly, 
this allows us to map a fixed instance $f_V(M)$ to a random one with the same $V$.  
It follows that, for any fixed $V$, if $f_V$ can be inverted on
even a $1/\poly(n)$ fraction of matrices $M$, then there is a
probabilistic algorithm that inverts it on arbitrary inputs $M$. 
A similar though more complicated assertion can be made about uniform
hardness with respect to choice of $V$ (see
Section~\ref{sec:hard-core}).  

Moreover, we show in Section~\ref{sec:hard-core} that 
reconstructing partial information about $f_V^{-1}(x)$ is almost as hard as inverting $f_V$.  
Specifically, assuming that $f_V$ is a one-way function, we show that \emph{any
entry of $M$ is hard to recover in any basis}, though this requires a
quasipolynomial hardness assumption on $f_V$.  We observe, also, that
$\tr \,M$, the trace of $M$, is hard to recover even under typical
super-polynomial hardness assumptions.

It remains an open question whether we can embed a trapdoor in $f_V$
or a suitable modification. We should point out that there are some
classical cryptosystems that are not known to be breakable by a quantum 
computer---lattice-based cryptosystems such as the
Ajtai-Dwork~\cite{Ajtai:1997:PKC} cryptosystem and their subsequent
improvements due to Regev~\cite{Regev:2004:NLB}, and the McEliece
cryptosystem~\cite{McEliece:Public}. Indeed, Regev's improvement in
the efficiency of lattice-based cryptosystems is based on a quantum
reduction---thus the increased efficiency is predicated on
resistance of the cryptosystem to quantum attacks!  Evidence of
quantum intractibility for this cryptosystem comes from the
relationship between finding short vectors and the dihedral hidden
subgroup problem~\cite{Regev:2004:QCL}. In particular, even though
single register Fourier sampling is information-theoretically
sufficient to reconstruct the hidden subgroup, the classical
reconstruction problem is as hard as Subset Sum. On the other hand,
quantum reconstruction is not ruled out, and Kuperberg's quantum
sieve~\cite{K} provides what may be thought of as a mildly
subexponential quantum reconstruction algorithm.  

The evidence for quantum intractibility for the one-way function proposed 
here is stronger: single register Fourier sampling is provably
insufficient, highly-entangled measurements on polynomially many registers is
necessary, and no Kuperberg-like approach can yield an efficient
algorithm. The design of efficient cryptographic primitives resistant
to quantum attack is a pressing practical problem whose solution can
have an enormous impact on the practice of cryptography long before a
quantum computer is physically realized.  A program to create such
primitives must necessarily rely on insights into the limits of
quantum algorithms, and this paper explores consequences of the
strongest such insights we have about the limits of quantum
algorithms.

\paragraph{Notation.} As above, we let $\F = \F_q$ denote the finite field with $q$ elements, $q$ a fixed prime. We let $\GL_n(\F_q)$ (abbreviated $\GL_n$ when the context is clear) denote the collection of invertible $n \times n$ matrices over $\F_q$. Similarly $\End_n = \End_n(\F_q)$ denotes the set of all $n \times n$ matrices. If $M \in \End_n$ and $V \subset \F_q^n$, we let $M V$ denote the collection $\{ M\vv \mid \vv \in V\}$.

\section{The function is one-to-one}
\label{sec:injective}

Our first theorem shows that when $m$ is slightly larger than $n$,
then $f_V$ is a one-to-one function with high probability.  We have
made only desultory attempts to optimize the rate at which $\delta =
m-n$ must grow for the theorem to hold.

\begin{theorem}
There is a constant $A$ such that if $m=n+\delta$ where $\delta \ge A
\ln^2 n$, then $f_V$ is one-to-one with high probability in $V$.
\end{theorem}

\begin{proof}  If there are two matrices $M, M'$ such that $MV=M'V$, 
then $KV=V$ where $K=M^{-1} M'$.  In other words, there is a
permutation $\pi \in S_m$ such that $K\vec{v}_i = \vec{v}_{\pi(i)}$ for all $i$.
We will show that with high probability $K=\one$ is the only matrix
with this property, and therefore that $M=M'$.

Let us call a particular permutation $\pi \in S_m$ \emph{consistent}
if there is a $K$ such that $K\vec{v}_i = \vec{v}_{\pi(i)}$ for all $i$, and let
$\cons_\pi$ be this event.  We will show that
\[ \Pr\left[ \bigvee_{\pi \ne 1} \cons_\pi \right] = o(1) \enspace. \]
i.e., with high probability the only consistent permutation is the identity $\pi=1$.

Given a fixed $\pi$, we determine an order on $V$ as follows.  First,
we sort the cycles of $\pi$ in order of increasing length, starting
with the fixed points.  We break ties by assigning each cycle an index
equal to the smallest $i$ such that $\vec{v}_i$ appears in it and putting
cycles with the smallest index first.  Then, we rotate each cycle so
that the $\vec{v}_i$ with smallest $i$ in that cycle comes first.  The
details here are irrelevant; all that matters is that each $\pi$
determines an order on $V$ with the properties that the vectors
corresponding to fixed points come first, and that groups of vectors
corresponding to cycles of $\pi$ are contiguous.

Now fix a constant $C$, and let $L_\pi$ consist of the first
$n+\delta-C \ln n$ vectors in $V$ according to this order.  Let
$\spans_\pi$ be the event that $L_\pi$ spans the entire space
$\F_q^n$.  Then the union bound gives
\begin{align*}
\Pr\left[ \bigvee_{\pi \ne 1} \cons_\pi \right] 
%&= \Pr\left[ \bigvee_{\pi \ne 1} \cons_\pi \wedge \bigwedge_\pi \spans_\pi \right] 
%+ \Pr\left[ \bigvee_{\pi \ne 1} \cons_\pi \wedge \overline{\spans_\pi} \right] \\
%&\le \Pr\left[ \bigvee_{\pi \ne 1} \cons_\pi \wedge \spans_\pi \right] 
%+ \Pr\left[ \bigvee_{\pi \ne 1} \overline{\spans_\pi} \right] \\
&\le \sum_{\pi \ne 1} \Pr\left[ \cons_\pi \!\mid\! \spans_\pi \right] 
+ \Pr\left[ \bigvee_{\pi} \overline{\spans_\pi} \right] \\
\end{align*}

To bound the conditional probability $\Pr[\cons_\pi \!\mid\!
\spans_\pi]$, note that if $L_\pi$ spans the entire space, then $K$ is
determined by the images of the vectors in $L_\pi$.  Therefore, if all
the vectors in $L_\pi$ are fixed by $K$, then $K=\one$ and $\pi=1$.
On the other hand, we have sorted $V$ so that the fixed vectors come
first, so if $\pi \ne 1$ none of the the $C \ln n$ vectors outside
$L_\pi$ can be fixed.  We expose these vectors in sorted order.  For
each $\vec{v}_i \notin L_\pi$ which is not the first in its cycle, the
probability that $\vec{v}_i$ is the image under $K$ of its predecessor
$\vec{v}_{\pi^{-1}(i)}$ is $q^{-n}$ since $\vec{v}_i$ is uniformly random.  These
events are independent and each of these cycles is of length at least
$2$, so the probability that $K \vec{v}_i = \vec{v}_{\pi(i)}$ for all $\vec{v}_i \notin
L$ is at most $q^{-(C/2) n \ln n}$.  Summing over all $(n+\delta)!$
permutations $\pi$ and assuming for simplicity that $\delta \le n$ (a
condition which we can easily remove), the conditional probability
that any $\pi \ne 1$ is consistent is at most
\[ (2n)! \,q^{-(C/2) n \ln n} = n^{O(1)} (2/\e)^{2n} n^{(2-(C/2) \ln q)n} \]
which is $o(1)$ if
\begin{equation}
\label{eq:c1}
C \ge 4/\ln q \enspace . 
\end{equation}

Now we bound the probability that $\spans_\pi$ fails to hold for any
$\pi$ by proving that with high probability $V$ contains no subsets
$L$ of size $n+\delta-C \ln n$ which do not span the entire space.  By
Markov's inequality, the probability that a given such $L$ does not
span the space is at most the expected number of nonzero vectors $\vec{u}$
which are perpendicular to all $\vec{v} \in L$.  Since the $\vec{v} \in V$ are
uniformly random, for any fixed $\vec{u}$ the inner product $\vec{u} \cdot \vec{v}$ is
zero with probability $1/q$.  Thus this expectation is
\[ 
(q^n-1)/q^{n+\delta-C \ln n} 
< q^{-\delta+C \ln n} 
= n^{O(1)} n^{-(A \ln q) \ln n}
\]
where we used $\delta = A \ln^2 n$.  The number of subsets of size $n+\delta-C \ln n$ is
\[ 
{n+\delta \choose C \ln n} 
< (2n)^{C \ln n} 
= n^{O(1)} n^{C \ln n}
\]
where we again assume for simplicity that $\delta \le n$.  So, by the union bound, the probability that a non-spanning subset of size $n+\delta-C \ln n$ is at most
$n^{O(1)} n^{(C-A \ln q) \ln n}$
which is $o(1)$ if
\begin{equation}
\label{eq:c2}
 A > C / \ln q \enspace .
\end{equation}

In order to satisfy~\eqref{eq:c1} and~\eqref{eq:c2}, we set, say, $C =
4/\ln q$ and $A = 5/\ln^2 q$.  Then with high probability, the
identity permutation $1$ is the only consistent one.  Finally, note
that $V$ spans the entire space with overwhelming probability; and in
this case, if $K\vec{v} = \vec{v}$ for all $\vec{v}$ in $V$, then $K$ must be the
identity.
\end{proof}

\section{Evidence for immunity against hidden subgroup attacks}
\label{sec:hardness}

In this section we relate the hardness of our function to several
fundamental problems in the area of quantum computation.  Our
principal hardness result, suggesting that $f_V$ can resist the
quantum attacks which Shor applied so dramatically to factoring and
discrete log, shows that Graph Isomorphism can be reduced to the
problem of inverting $f_V$.  Our current belief, based on a series of
negative results, is that Graph Isomorphism, and more generally the
HSP on groups like $S_n$ and $\GL_n$ which have exponentially
high-dimensional representations, is hard for quantum computers.  If
this belief is correct, then $f_V$ cannot be efficiently inverted by
such methods. We observe, also, that inverting $f_V$ can be reduced to
natural hidden shift and hidden subgroup problems on the group
$\GL_n$.

We begin by reducing the problem of inverting $f_V$ to the Hidden
Shift Problem on the group $\GL_n$.  Given a group $G$, an instance of
a Hidden Shift problem consists of two functions $f_1, f_2:G \to S$,
with the promise that $f_2(g) = f_1(gs)$ for some shift $s \in G$.
Now, given $V$ and $f_V(M) = MV$, we can define two functions $f_1,
f_2: \GL_n \to S$ where $S$ is the set of unordered lists of vectors
in $\F_q^n$.  Namely, we define
\[ f_1(N) = NV \; \mbox{and} \; f_2(N) = N f_V(M) = NMV \enspace . \]
Then $f_1(N) = f_V(N)$ and $f_2(N) = f_V(NM) = f_1(NM)$, and $M$ is
the hidden shift.

Now, given a Hidden Shift Problem on a group $G$ where the functions
$f_1,f_2$ are one-to-one, we can reduce it to a Hidden Subgroup
Problem on a larger group, namely the wreath product $G \wr \Z_2$.
This group is the semidirect product $(G \times G) \rtimes \Z_2$,
where we extend $G \times G$ with an involution which exchanges the
two copies of $G$.  We denote its elements $(g_1,g_2,z)$, where those
with $z=0$ form the normal subgroup which fixes the two copies of $G$,
and those with $z=1$ form its nontrivial coset which exchanges them.

Recall that an instance of the Hidden Subgroup Problem consists of a
function $f:G \to S$ with the promise that, for some subgroup $H$,
$f(x)=f(y)$ if and only if $x=yh$ for some $h \in H$.  Given a Hidden
Shift Problem with functions $f_1,f_2:G \to S$, define the following
function $f:G \wr \Z_2 \to S^2$:
\begin{align*}
 f(g_1,g_2,0) &= (f_1(g_1), f_2(g_2)) \\
 f(g_1,g_2,1) &= (f_2(g_2), f_1(g_1)) 
\end{align*}
Now suppose that $f_2(g)=f_1(gs)$ and let $\alpha$ be the involution
$(s^{-1},s,1)$.  If multiplication in $G \wr \Z_2$ is defined so that
$(g_1,g_2,0)\cdot \alpha = (g_2 s, g_1 s^{-1}, 1)$, then $f$'s hidden
subgroup is the order-2 subgroup $H=\{1,\alpha\}$.  (Indeed, the
canonical reduction of Graph Isomorphism to the Hidden Subgroup
Problem over $S_n \wr \Z_2$ is exactly of this type, where
$\alpha=(\pi^{-1},\pi,1)$ exchanges the two graphs and $\pi$ is the
isomorphism between them.)  Finally, we point out that $\GL_{2n}$
contains a copy of $\GL_n \wr \Z_2$: namely, the subgroup consisting
of matrices of the form
\[ 
\begin{pmatrix} g_1 & 0 \\ 0 & g_2 \end{pmatrix}
\; \mbox{or} \;
\begin{pmatrix} 0 & g_1 \\ g_2 & 0 \end{pmatrix}
\]
where $g_1, g_2 \in \GL_n$.  Thus the problem of inverting $f_V$ reduces to the Hidden Shift and Hidden Subgroup Problems in $\GL_n$ and $\GL_{2n}$ respectively. 

Now, we give a reduction from Graph Isomorphism to the problem of
inverting $f_V$.  Specifically, we reduce the decision problem of
telling whether two graphs $G_1, G_2$ are isomorphic to the decision
problem of telling, given $V$ and $W$, whether there is a matrix $M$
such that $MV=W$, and hence whether $W$ is in the image of $f_V$.  The
same construction reduces the promise problem of finding the
isomorphism between two isomorphic graphs to the problem of finding
$M=f_V^{-1}(W)$.

The reduction is quite simple.  Given a graph $G_1$ with $n$ vertices
and $m$ edges, $V$ will consist of $n+m$ vectors in $\F_q^n$.  We
identify each vertex $u$ with a basis vector $\vec{u}$, which we include in $V$,
and for each edge $(u,v)$ we include the vector $\vec{u}+\vec{v}$.  We construct
$W$ from $G_2$ similarly.

Clearly $G_1 \cong G_2$ if and only if $MV=W$ for some permutation
matrix $M$.  First we show that, if $q \ge 3$, any $M$ such that
$MV=W$ is necessarily a permutation matrix.  To see this, note that
since each vertex of $G_1$ gets mapped to a vertex or an edge of
$G_2$, each column of $M$ is zero except for one or two $1$s.  But in
$\F_q^n$ with $q \ge 3$, the sum of two such vectors has at least two
nonzero components, so no edge of $G_1$ can be mapped to a vertex of
$G_2$.  It follows that every vertex of $G_1$ is mapped to a vertex of
$G_2$, so $M$ is a permutation matrix.

In the case $q=2$, it is possible that $M$ is not a permutation
matrix, and that some vertices get mapped to edges and vice versa.
However, $M$'s existence still implies that $G_1$ and $G_2$ are
isomorphic, and allows us to easily determine the isomorphism $\pi$
between them.
Let us call a vertex of $G_1$ ``green'' or ``red'' if it is mapped to
a vertex or an edge, respectively, and consider a vertex $w$ of $G_2$.
Since $M^{-1}w$ is either a vertex or an edge, either there is a green
vertex $u$ such that $M\vec{u} = \vec{w}$, or there is a red vertex $u$ with a
unique green neighbor $v$ such that $M\vec{u}=\vec{w}+M\vec{v}$ and so $M(\vec{u}+\vec{v})= \vec{w}$.  In
either case, define $\pi(u)=w$; since $\pi$ is one-to-one, it follows
that \emph{every} red vertex has a unique green neighbor.

It remains to check that $\pi$ is an isomorphism.  Denote the set of
edges of $G_1$ and $G_2$ as $E_1$ and $E_2$ respectively, and suppose
that $(u,v) \in E_1$.  If $u$ and $v$ are green, then
$M(\vec{u}+\vec{v})=\vec{\pi(u)}+\vec{\pi(v)}$.  If $u$ is red and $v$ is its unique green
neighbor, then $M\vec{u}=\vec{\pi(u)}+\vec{\pi(v)}$.  Finally, if $u$ and $v$ are both
red, they must have the same green neighbor $t$ since otherwise
$M(\vec{u}+\vec{v})$ would be the sum of four basis vectors; then $M(\vec{u}+\vec{v}) =
\vec{\pi(u)}+\vec{\pi(v)}+2 \vec{\pi(t)} = \vec{\pi(u)}+\vec{\pi(v)}$.  In each case, since
$\vec{\pi(u)}+\vec{\pi(v)} \in W$ we have $(\pi(u),\pi(v)) \in E_2$, and this
completes the proof.

\section{Uniformity of hardness, amplification, and hard-core predicates}
\label{sec:hard-core}

\paragraph{Self-reducibility and uniform hardness.}  

As we pointed out in the Introduction, our function has a simple symmetry 
which causes it to be self-reducible from the worst case to the random case: 
for any fixed $V$, we have $f_V(AM) = A f_V(M)$.  It follows by standard 
amplification that, for any fixed $V$, if $f_V$ can be inverted on even a $1/\poly(n)$ fraction of
matrices $M$ then it can be inverted with probability $1 - e^{-\poly(n)}$ on any particular $M$.  

We can define uniform hardness with respect to $V$ using another obvious symmetry,
\[ f_{BV}(M) = f_V(MB) \enspace . \]
Let us say that $V \sim V'$ if there is a $B \in \GL_n$ such that $V'=BV$.  This is clearly an 
equivalence relation; we will call the equivalence class containing $V$ its \emph{orbit}, 
and denote it $[V]$.  Then a similar argument shows that inverting $f_V$ is uniformly hard 
within each orbit: namely, if $f_V$ 
can be inverted on even a $1/\poly(n)$ fraction of
matrices $M$ and vectors  $V' \in [V]$ then it can be 
inverted with probability $1 - e^{-\poly(n)}$ on any particular 
$M$ and $V' \in [V]$.

\remove{
We begin by pointing 
out a simple self-reducibility property of the function $f_V$ which
will be applied in the hardness proofs that follow: for invertible
matrices $A, B \in \GL_n(\F)$,
$$
f_{BV}(AM) = A f_V(MB) = [AMB]V\enspace.
$$ 
It follows, for example, by standard amplification that if the
function $f_V$ can be inverted on even a $1/\poly(n)$ fraction of
matrices $M$ then the function can be inverted with probability $1 -
e^{-\poly(n)}$ on any particular $M$.  Similarly, the mapping from $V$
to $BV$ partitions the vectors $V$ into orbits, within each of which
the function $f_V$ is uniformly hard.

To formulate uniform hardness with respect to $V$,
define $V \sim V'$, for two sets of vectors of $\F^n$, if
there is an element $T$ of $\GL_n$ for which $V = TV'$. It is easy to
see that $\sim$ is an equivalence relation; we let $[V]$ denote the
equivalence class of $V$, which we call the \emph{orbit} of
$V$. Now it is easy to see that the function $f_V$ 
can be inverted on even a $1/\poly(n)$ fraction of
matrices $M$ and vectors  $V' \in [V]$ then the function can be 
inverted with probability $1 - e^{-\poly(n)}$ on any particular 
$M$ and $V'$. 
}

A priori, even if it is hard to invert $f_V$, one might hope to recover 
partial information about $M$ from its image $f_V(M)$, such as its trace 
or a single entry in some basis.  In this section, we show that this is essentially as hard 
as recovering all of $M$.  Therefore, under reasonable hardness assumptions regarding $f_V$, 
these goals are also impossible for quantum computers to carry out efficiently.

\paragraph{Hard-core predicates.}

A hard-core predicate is an efficient description of a bit of information that is concealed by a given one-way function. Specifically, if $\{ f_n : D_n  \rightarrow R_n \}$ is a family of one-way functions, then an \emph{$s(n)$-hard-core predicate} is a polynomial time computable family of functions $\{ b_n : D_n \rightarrow \{0,1\}\}$ so that for any algorithm $A$ running in time $s(n)$, for sufficiently large $n$,
$$
\left| \Pr_{f_n,w}[A(f_n(w)) = b_n(w)] - \frac{1}{2} \right| \leq \frac{1}{s(n)}\enspace.
$$
Our goal here is to show that every individual entry of $M$ is a hard-core bit in any basis; in particular, \emph{recovering any entry of $M$ is as hard as inverting $f_V$}. We also point out that recovering the trace of $M$ is as hard as inverting $f_V$.

We begin by formalizing the notions of hardness we require for the function $f_V$.
\begin{assumption}[$t(n)$-hardness]\label{assume:poly-hardness}
For each $n \geq 1$, let $m = m(n) = (1+ \epsilon)n$ for some constant $\epsilon > 0$, let $M$ be a uniformly random element of $\GL_n(\F)$, and let $V$ be a collection of $m$ independently and uniformly selected elements of $\F^n$. Then for all quantum algorithms $A$ running in time $t(n)$,
\[
\Pr_{V,M}[A(M(V), V) = M] = \frac{1}{t(n)}\enspace.
\]
\end{assumption}

We devote the remainder of this section to showing the following two theorems.

\begin{theorem}
If $f_V$ is quasipolynomially hard (that is, $t(n)$-hard for every $t(n) = 2^{\log^{O(1)} n}$) then every entry of $M$ (in any basis) is a quasipolynomially hard-core predicate.
\end{theorem}

\begin{theorem}
If $f_V$ is polynomially hard (that is, $t(n)$-hard for every $t(n) = n^{O(1)}$) then the trace $\tr: \GL_n(\F) \rightarrow \F$ is a polynomially hard-core predicate.
\end{theorem}

\subsection{The bilinear predicate: every matrix entry is hard}

Given two basis vectors $\va$ and $\vb$, the corresponding matrix element 
can be written as an inner product $\langle \va, M\vb\rangle$.  
We will show that if $f_V$ is quasipolynomially hard,
then this function is a hard-core predicate for $f_V$ 
for any fixed nonzero $\va, \vb \in \F^n$.  
Specifically, given an algorithm $P$ running in time $2^{\log^{O(1)} n}$ 
for which 
\[ 
\Pr_{V,M} \bigl[ P(f_V(M), V) = \langle a, Mb\rangle \bigr] \geq 1/2 + \epsilon 
\;\mbox{ with }\;
\epsilon = 2^{-\log^{O(1)} n}
\enspace ,
\] 
we show how to invert $f_V$ on a $2^{-\log^{O(1)} n}$ fraction of its
inputs $M$, which would contradicting the assumption that $f_V$ is quasipolynomially hard. 

To simplify the exposition, we will fix $q$ to be $2$ in this section, 
and write $\F=\F_2$.  We rely on the Goldreich-Levin theorem~\cite{Goldreich:1989:HCP}; 
for larger prime $q$, we rely on its generalization to arbitrary finite fields by 
Goldreich, Rubinfeld, and Sudan~\cite{GoldreichRS1995}.

Initially, we wish to focus attention on certain ``good'' choices of $V$, where
the algorithm $P$ is a good predictor for $\langle \va, M \vb \rangle$. 
\iffalse
In
particular, define $V \sim V'$, for two sets of vectors of $\F^n$, if
there is an element $T$ of $\GL_n$ for which $V = TV'$. It is easy to
see that $\sim$ is an equivalence relation; we let $[V]$ denote the
equivalence class of $V$, which we call the \emph{orbit} of
$V$. 
\fi
Recall that $[V]$ denotes the orbit of $V$ under multiplication
by elements of $\GL_n$.
Define an element $V$ to be ``good'' if
\begin{equation}
\label{eq:good}
\Pr_{V' \in [V],M}\bigl[ P(f_{V'}(M), V') = \langle a, Mb\rangle  \bigr] \geq \frac{1}{2} + \frac{\epsilon}{2}\enspace.
\end{equation}
It is easy to show that at least an $\epsilon/2$ fraction of $V$ must be good in this sense; we fix a specific such $V$ for the remainder of the proof, and show how to invert the function $f_V$ in this case.

We first show how to use $P$ to implement an algorithm for any fixed $M$,
which takes as input $x,y \in \F^n$ (and $(f_{V} (M), V)$) and outputs
$\langle x, My\rangle$ correctly on $1/2+ \epsilon/2$ fraction of
$x,y$.  First note that for two matrices $A, B \in \GL_n$, 
the pair $(f_{BV}(AMB^{-1}), BV) = (AMV, BV)$ can be computed efficiently from
$(f_{V}(M), V)=(MV,V)$ by left-multiplying $MV$ and $V$ by $A$ and
$B$ respectively.  Defining $T(A, B) = P(f_{BV}(AMB^{-1}), BV)$, we may then 
rewrite~\eqref{eq:good} in terms of $T(\cdot,\cdot)$:
\begin{equation}
	\label{eq:good-matrix}
	\Pr_{A, B \in \GL_n(\F)} [T(A, B) = \langle a, AMB^{-1}b\rangle] \geq \frac{1}{2} + \frac{\epsilon}{2}\enspace.
\end{equation}
Finally, for a pair of vectors $\vx, \vy \in \F^n$, define $t(\vx, \vy) = T(A, B)$, where $A$ and $B$ are random elements of $\GL_n(\F)$ for which $A^t \va = \vx$ and $B^{-1} \vb = \vy$, so that $\langle \va, AMB^{-1} \vb\rangle = \langle \vx, M \vy \rangle$. Rewriting~\eqref{eq:good-matrix}, we conclude:
\begin{equation}
	\Pr_{\vx, \vy \in \F^n} [t(\vx, \vy) = \langle \vx, M \vy \rangle] \geq \frac{1}{2} + \frac{\epsilon}{2}\enspace.
\end{equation}
Let us call a vector $\vx \in \F^n$ \emph{$\ell$-good} if 
$\Pr_{\vy \in \F^n} [t(\vx, \vy) = \langle \vx, M \vy\rangle] \geq 1/2 + \epsilon/4$. If
follows that a uniformly selected $\vx$ is $\ell$-good with probability
at least $\epsilon/4$. Note, furthermore, that if $\vx$ is a fixed
$\ell$-good element of $\F^n$, then the Goldreich-Levin construction~\cite{Goldreich:1989:HCP} 
can be used to determine $\langle \vx, M \vy \rangle$ for all $\vy \in \F^n$
(in time polynomial in $n$ and $\epsilon^{-1}$). In particular, this
determines an entire row of $M$ when expressed in a basis containing $\vx$.

We consider now a family $G$, consisting of $2 \log m$ vectors
selected independently and uniformly in $\F^n$. We say that $G$ is
$\ell$-good if this is true of each of its elements, a favorable event
that occurs with probability at least $(\epsilon/4)^{\log 2m}$. 
Furthermore, the probability that $G$ contains a linearly
dependent set of vectors is no more than $2 \log(m) \cdot 2^{-n +
2\log m} = 2^{-\Omega(n)}$. (This can be seen by selecting the
elements of $G$ in order and bounding the unlikely event that an
element falls into the span of the previously chosen vectors.) Thus
\[ 
\Pr[ G~\text{is $\ell$-good} \wedge G~\text{is independent}] 
\geq (\epsilon/4)^{2\log m} + e^{-\Omega(n)}
\enspace . 
\] 
Now, for each $\vg \in G$, application of the Goldreich-Levin construction to each component of $\vg$
(reconstructing $\langle \vg, M \vy\rangle$ for all $\vy$) determines
$\langle \vg, M \vv \rangle$ for each $\vv \in V$ and $\vg \in G$.  Therefore, 
in this case we can reconstruct $2 \log m$ ``generalized rows'' of $M$. 

Observe that if the elements of $V$ (and hence $W=M(V)$) are
considered to be selected independently and uniformly at random, 
{\em and} independently of $G$, then
the probability that two elements $w$ and $w'$ of $W$ have the
property that $\langle \vg, \vw \rangle = \langle \vg, \vw' \rangle$ for all
$\vg \in G$ is $2^{-2 \log m}$. Let $\Pi_G: \F^n \rightarrow \F^{2 \log m}$ 
denote the projection onto the space spanned by the vectors in
$G$. In particular, this information would appear to determine the
bijection $b_M: V \rightarrow W$ effected by the action of $M$ on
$V$. This intuitive argument is misleading, as written, since the
notion of $\ell$-good depends on $V$ (and so on $W$) via the 
arbitrary predicting algorithm $P$. Instead, our
goal below will be to show that the total number of permutations of
the set $W$ under which $\Pi_G$ is invariant is small enough that we
can exhaustively search them to uncover the bijection $b_M$ and hence
the linear operator $M$.

Consider random (and independent) selection of $G$, $V$, and $M$ (so
that $W = M(V)$ is also determined) with no extra conditioning except
that $G$ be linearly independent. Let $I_G$ denote the collection of
permutations $\phi: M \rightarrow M$ with the property that $\Pi_G w =
\Pi_G \phi(w)$, for all $w \in W$. We will show below that  
$\Exp_{V,M,G}[|I_G|] = O(\sqrt{m})$. Then Markov's inequality
will allow us to bound the probability that $|I_G|$ exceeds 
$\epsilon^{O(\log n)}$. To round out the proof we will show 
that the chance that $V$ is good and that $G$ is $l$-good is 
much higher than this failure probability, thereby concluding that
there is a significant chance that $V$ is good, $G$ is 
$l$-good and that $|I_G| = \epsilon^{O(\log n)}$. 

As the elements of $w$ are selected
independently (and uniformly) in $\F^n$, each $\Pi_G w$ is an
independent, uniform element of $\F^{|G|}$. Fixing a permutation
$\phi$, let $\lambda_1, \lambda_2, \ldots$ be the lengths of its
cycles, arranged in nonincreasing order. The probability that the
elements of $M$ in each of these cycles are mapped to the same element
under $\Pi_G$ is no more than $\prod_i (2^{-|G|})^{\lambda_i -1} =
\prod_i (m^{-2})^\tau(\phi)$, where $\tau(\phi) = \sum_i (\lambda_i -
1)$ is also the minimum number of transpositions required to write
$\phi$.

This quantity is bounded by the lemma below. Its proof uses the machinery of exponential generating functions, and is relegated to Appendix~\ref{app:egf}.
\begin{lemma}
\label{lem:egf}
Let $0 < z < 1/k$; then 
\begin{equation}
\label{eq:qn}
 q_k(z) = \sum_{\pi \in S_k} z^{t(\pi)}  = O(\sqrt{k}) \,\frac{e^{-k}}{(1-zk)^{1/z}} \enspace .
\end{equation}
\end{lemma}

In light of this bound, the expectation of $|I_G|$, the number of $\phi$ under which $\pi_G$ is invariant, is no more than
$$
\sum_{\phi \in S_m} \left(\frac{1}{m^2}\right)^{\tau(\phi)} = O(\sqrt{m}) \frac{e^{-m}}{(1 - 1/m)^{m^2}}\enspace.
$$
As $- \ln(1 - x) = x + x^2/2 + x^3/3 + \ldots$, we have
$$
e^{-m} \cdot (1 - 1/m)^{-m^2} = \exp(-m + m^2[ 1/m + (1/m)^2/2 + O(1/m^3)]) = O(1)\enspace.
$$
Thus $\Exp[|I_G|] = O(\sqrt{m})$.

Putting the pieces together, with $M$, $V$, and $G$ selected as above,
$$
\Pr_{V, M, G}  [ (\text{$V$ is good}) \wedge (\text{$G$ is both $\ell$-good and linearly independent})] \geq \frac{\epsilon}{2} \cdot \left(\frac{\epsilon}{4}\right)^{2 \log m} \geq \left(\frac{\epsilon}{4}\right)^{1 + 2 \log m}\enspace.
$$
As $\Exp_{V,M,G}[|I_G|] = O(\sqrt{m})$, by Markov's inequality there is a constant $c$ so that
$$
\Pr_{V,M,G}\left[|I_G| \geq  c \sqrt{m} (4/\epsilon)^{2 \log m}\right] \leq \frac{1}{2} \cdot \left[\left(\frac{\epsilon}{4}\right)^{1 + 2 \log m}\right]\enspace.
$$
Thus, with probability at least $(1/2) (\epsilon/4)^{(1 + 2 \log m)}$, $V$ is good, $G$ is $\ell$-good, and there are $(4/\epsilon)^{O(\log n)}$ permutations of $W$ that fix $\Pi_G$. These permutations determine a set of no more than $(4/\epsilon)^{O(\log n)}$ mappings between $V$ and $W$ consistent with $M$; these can be exhaustively searched in time $\text{poly}(n) \cdot (\epsilon/4)^{O(\log n)}$, which is quasipolynomial when $\epsilon^{-1}$ is.

\smallskip
We conclude this section with a proof that, even if $\epsilon^{-1}$ is only polynomial in $n$, hardness with respect to quasipolynomial time is the most we can hope for in the case of the bilinear predicate (in absence of further information about the preimage).  First, choose a subspace $S$ of $\F^n$ with dimension $\dim S = \log_2 n$.  Now consider an oracle $P(a,b)$ defined as follows.  If either $a$ or $b$ is orthogonal to $S$, then $P(a,b) = \langle a, Mb \rangle$, but if neither of them is orthogonal to $S$, then $P(a,b)$ is uniform in $\F$.  Since a uniform vector in $\F^n$ is orthogonal to $S$ with probability $1/n$, it follows that $P(a,b)$ is correct with probability $1/2 + \epsilon$ where $\epsilon > 1/n$. 

Now choose a basis for $\F^n$, and let $S$ be the subspace generated by the first $\dim S$ basis vectors.  It is clear that this oracle gives us no information whatever regarding the matrix elements in the $\dim S \times \dim S$ minor at the upper left-hand corner of $M$.  Therefore, we are forced to try all possible values for the elements of this minor by exhaustive search, and this takes $2^{(\dim S)^2} = 2^{\log^2 n}$ time.

\subsection{The trace predicate}

The proof that the trace predicate is hard is a direct consequence of the Goldreich-Levin theorem~\cite{Goldreich:1989:HCP} and its generalization to arbitrary finite fields by Goldreich, Rubinfeld, and Sudan~\cite{GoldreichRS1995}. Specifically, consider the trace $\tr: \GL_n(\F) \rightarrow \F$. Suppose now that there is a polynomial-time quantum algorithm $P$ so that for $M$ selected uniformly at random in $\GL_n$ and $V$ a collection of $m$ independent and uniform vectors of $\F^n$, 
\[
\Pr_{V,M} \bigl[ P(f_V(M),V) = \tr(M) \bigr] \geq \frac{1}{2} + \epsilon\enspace,
\]
where $\epsilon = n^{-O(1)}$. It follows that for at least an $\epsilon/2$ fraction of the $V$, when selected as above, we have
\[
\Pr_{M} \bigl[ P(f_V(M),V) = \tr(M) \bigr] \geq \frac{1}{2} + \frac{\epsilon}{2}\enspace.
\]
We show how to invert $f_V$ for such ``good'' $V$; as these occur with probability $\epsilon/2$, this would contradict the assumption that $f_V$ is polynomially hard. For the remainder of the proof we fix a specific $V$ satisfying the the equation above.

Again note that for any matrix $N \in \GL_n$, the collection $f_V(NM)=NMV$ can be computed in polynomial time from $f_V(M)$, simply by left-multiplying the collection $f_V(M)=MV$ by $N$.  In particular, given $f_V(M)$, the function $T: \GL_n(\F) \rightarrow \F$ given by $T(N) = P(f_V(NM),V)$ can be computed in polynomial time and has the property that
\begin{equation}
\label{eq:over-GL}
\Pr_N \bigl[ T(N) = \tr(NM) \bigr]  \geq \frac{1}{2} + \epsilon\enspace.
\end{equation}
Now, for a fixed matrix $C$, the function $\ell_C: M \mapsto \tr(CM)$ is a linear function and, moreover, all linear combinations of the entries of $M$ 
can be written in this way. In light of this, note that if the guarantee~\eqref{eq:over-GL} could be arranged with the matrix $C$ being selected uniformly at random from the collection of \emph{all} matrices (rather than the invertible ones), we could immediately apply the Goldreich-Levin~\cite{Goldreich:1989:HCP} construction at this point to recover $M$. This ``oracle'' $T$ can, however, be extended to an oracle $\tilde{T}$ defined on the family of all matrices $C$ by simply assigning random values to the singular matrices $C \not\in \GL_n$, in which case with constant probability (over the selection of random values for this oracle),
\begin{equation}
\label{eq:over-End}
\Pr_N \bigl[ T(N) = \tr(NM) \bigr]  \geq \frac{1}{2} + \alpha_p(n) \epsilon\enspace,
\end{equation}
where 
\[
\alpha_p(n) = \prod_{i = 0}^{n-1} \left( 1 - \frac{1}{p^{n-i}}\right) \geq \prod_{i = 0}^{\infty} \left( 1 - \frac{1}{2^{i}}\right) \approx .2711
\]
is the probability that a random $n \times n$ matrix over $\F_p$ is invertible. In this case, when $p = 2$ the Goldreich-Levin theorem can be applied directly:

\begin{theorem}[\cite{Goldreich:1989:HCP}]
Let $g: \F_2^n \rightarrow \F_2$ be a function so that for some $h \in \F_2^n$, $\Pr_{x \in \F_2^n} \left[ g(x) = \langle x, h \rangle \right] \geq \frac{1}{2} + \epsilon$ and let $c \geq 0$. Then there is a randomized algorithm running in time $\text{poly}(n, \epsilon^{-1})$ (and making no more than $\text{poly}(n, \epsilon^{-1})$ black-box queries to $g$) that determines $h$ with probability $1 - 1/n^c$.
\end{theorem}

\noindent 
When $q > 2$, one has to apply the generalization of~\cite{Goldreich:1989:HCP} to arbitrary finite fields by Goldreich, Rubinfeld, and Sudan~\cite{GoldreichRS1995}.

%\bibliography{BibTeX/stoc1980,BibTeX/stoc1990,BibTeX/stoc2000,BibTeX/math,BibTex/focs1990,BibTex/focs2000,BibTeX/siamjcomput,BibTeX/quantph,BibTeX/jacm,BibTeX/crypto}
\newcommand{\etalchar}[1]{$^{#1}$}
\ifx \k \undefined \let \k = \c \immediate\write16{Ogonek accent unavailable:
  replaced by cedilla} \fi\ifx \ocirc \undefined \def \ocirc
  #1{{\accent'27#1}}\fi\ifx \mathbb \undefined \def \mathbb #1{{\bf #1}}
  \fi\ifx \mathbb \undefined \def \mathbb #1{{\bf #1}}\fi\input
  path.sty\hyphenation{ Cher-vo-nen-kis Eh-ren-feucht Hal-pern Jean-ette
  Kam-eda Leigh-ton Mehl-horn Metro-po-lis Pra-sad Prep-a-ra-ta Press-er
  Pros-ku-row-ski Ros-en-krantz Ru-dolph Schie-ber Schnei-der Te-zu-ka
  Vis-wa-na-than Yech-ez-kel Yech-i-ali data-base data-bases dead-lock
  poly-adic }

\pagebreak

\appendix
\section{Proof of Lemma~\ref{lem:egf}}
\label{app:egf}

\begin{figure}
\label{fig:proof-details}
\begin{center}
\includegraphics[width=4in]{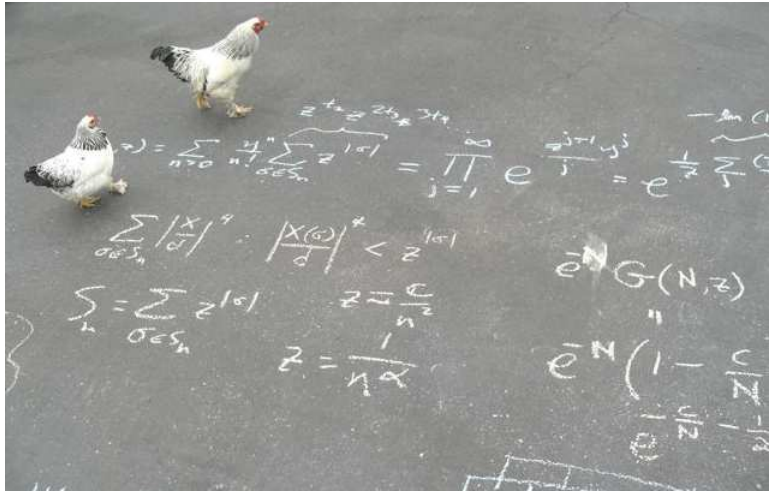}
\end{center}
\caption{Two of the authors hard at work chalking up the proof of Lemma~\ref{lem:egf} on an asphalt driveway.}
\vspace{2mm}
\hrule
\end{figure}

Recall that Lemma~\ref{lem:egf} asserts that if  $0 < z < 1/k$; then 
\begin{equation}
\label{eq:qn-recap}
 q_k(z) = \sum_{\pi \in S_k} z^{t(\pi)}  = O(\sqrt{k}) \,\frac{e^{-k}}{(1-zk)^{1/z}} \enspace .
\end{equation}

\begin{proof}[Proof of Lemma~\ref{lem:egf}]
Consider the exponential generating function 
\[ g(y,z) = \sum_{m=0}^\infty
\frac{y^m}{m!} \,q_m(z) \enspace . \]
Using the techniques of~\cite[Chapter 3]{Wilf94}, we can write this as a product
over all $k$ of contributions from the $(k-1)!$ possible $k$-cycles, including
fixed points.  Since each such cycle contributes $k$ to $m$ and $k-1$ to
$t(\pi)$, and since there are $(k-1)!$ $k$-cycles on a given set of $k$ objects,
it follows (cf. Figure~\ref{fig:proof-details}) that
\[ 
g(y,z) = \prod_{k=1}^\infty \exp\!\left( \frac{y^k z^{k-1}}{k} \right)
= \exp\!\left( \sum_{k=1}^\infty \frac{y^k z^{k-1}}{k} \right)
= \exp\!\left( -\frac{1}{z} \ln (1-yz) \right)
= \frac{1}{(1-yz)^{1/z}} \enspace . 
\] 
Now note that $e^{-k} g(k,z)$ is the expectation of $q_m(z)$, where $m$ is
Poisson-distributed with mean $k$.  Since $q_m(z) > 0$, this expectation is at
least $q_k(z)$ times the probability that $m=k$, which is $e^{-k} k^k / k! =
(1-o(1)) / \sqrt{2 \pi k}$.  Thus we have \[ q_k(z) \le (1+o(1)) \sqrt{2 \pi k}
\cdot e^{-k} g(k,z) \]
which concludes the proof. 
\end{proof}

\end{document}